\documentclass{PoS}

\usepackage{subcaption}
\usepackage{url}
\usepackage{cleveref}
\usepackage{multicol}
\usepackage{gensymb}
\usepackage{graphicx}
\usepackage{multirow}

\newcommand{\plm}{$\pm$}

\newcommand{\veritas}{{VERITAS}}

\newcommand{\fermi}{\textit{Fermi}}

\newcommand{\LAT}{\textit{Fermi}-LAT}

\newcommand{\swift}{\textit{Swift}}

\newcommand{\psr}{PSR J2032+4127}
\newcommand{\TeVJ}{TeV J2032+4130}
\newcommand{\mt}{MT91 213}

\newcommand{\fermipy}{\emph{fermipy}}

\newcommand{\GR}{$\gamma$-ray}

\newcommand{\ra}{$\alpha_{J2000}$}
\newcommand{\dec}{$\delta_{J2000}$}

\newcommand{\hms}[3]{$#1^\mathrm{h}\allowbreak#2^\mathrm{m}\allowbreak#3^\mathrm{s}$}

\newcommand{\dms}[3]{$#1\degree\allowbreak#2'\allowbreak#3''$}

\title{\psr, the counterpart of \TeVJ? Multiwavelength Monitoring of the Approach to Periastron}

\ShortTitle{\psr, the counterpart of \TeVJ?}

\author{\speaker{Ralph Bird}, for the \veritas\ Collaboration\thanks{http://veritas.sao.arizona.edu}\\
        University of California, Los Angeles\\
        E-mail: \email{ralphbird@astro.ucla.edu}}

\abstract{\psr\ has recently been identified as being in a long period (45-50 years) binary in a highly eccentric orbit with the Be star \mt. 
Periastron is due to occur in November 2017 and this rare occurrence has prompted a multiwavelength monitoring campaign to determine if the system is a \GR\ binary, and, if so, to study what would be only the second \GR\ binary with a known compact object. 
In the same direction as \TeVJ, \GR\ emission from this binary system could be related to the extended very high energy \GR\ emission from that region.
As part of this monitoring, observations are being conducted by \swift, \LAT\ and \veritas. 
We present the status of those observations, preliminary results and the plan for continued monitoring through periastron.}

\FullConference{35th International Cosmic Ray Conference\\
		 10-20 July, 2017\\
		 Bexco, Busan, Korea}

\begin{document}

\section{Introduction}
\TeVJ\ is an extended, very high energy (VHE, E $>$ 100~GeV) source lying in the direction of the Cygnus OB2 region \cite{HEGRA_TeV2032}.  
It was the first unidentified VHE \GR\ source and after thirteen years of observations it remains a mystery.
The most promising potential association is with the pulsar wind nebula of \psr\ as has been suggested in \cite{VTS_TeV2032}.

\psr\ has recently been identified as the compact object in a binary system with the Be star MT91~213 \cite{Lyne}.
This system has a highly eccentric orbit with a period of 45-50 years \cite{Ho}.
Periastron is due to occur in November 2017 and a multiwavelength campaign is underway to monitor this rare event.
These observations will be used to test whether \psr/\mt\ form a \GR\ binary\footnote{Here we define \GR\ binaries as binary systems containing a massive star and a compact object which have peak $\nu F_{\nu}\; > \; $1~MeV as in \cite{DUBUS}} and potentially solve the mystery of the origin of \TeVJ.

\psr\ was first identified by \LAT\ with a period of 142~ms \cite{LAT_PSR}.
Radio observations have since detected dramatic changes in its spin down rate, at a level never seen in an isolated pulsar, from $\dot{\nu}$ = -1$\times$10$^{-12}$~s$^{-1}$ to -2$\times$10$^{-12}$~s$^{-1}$ over the course of six years \cite{Lyne}.
This has been attributed to Doppler shifting from the motion of the pulsar within a binary system.
The binary companion has been identified as the 15~M$_{\odot}$ Be star \mt.

\begin{figure}[b]
\centering
  \includegraphics[width=0.4\textwidth]{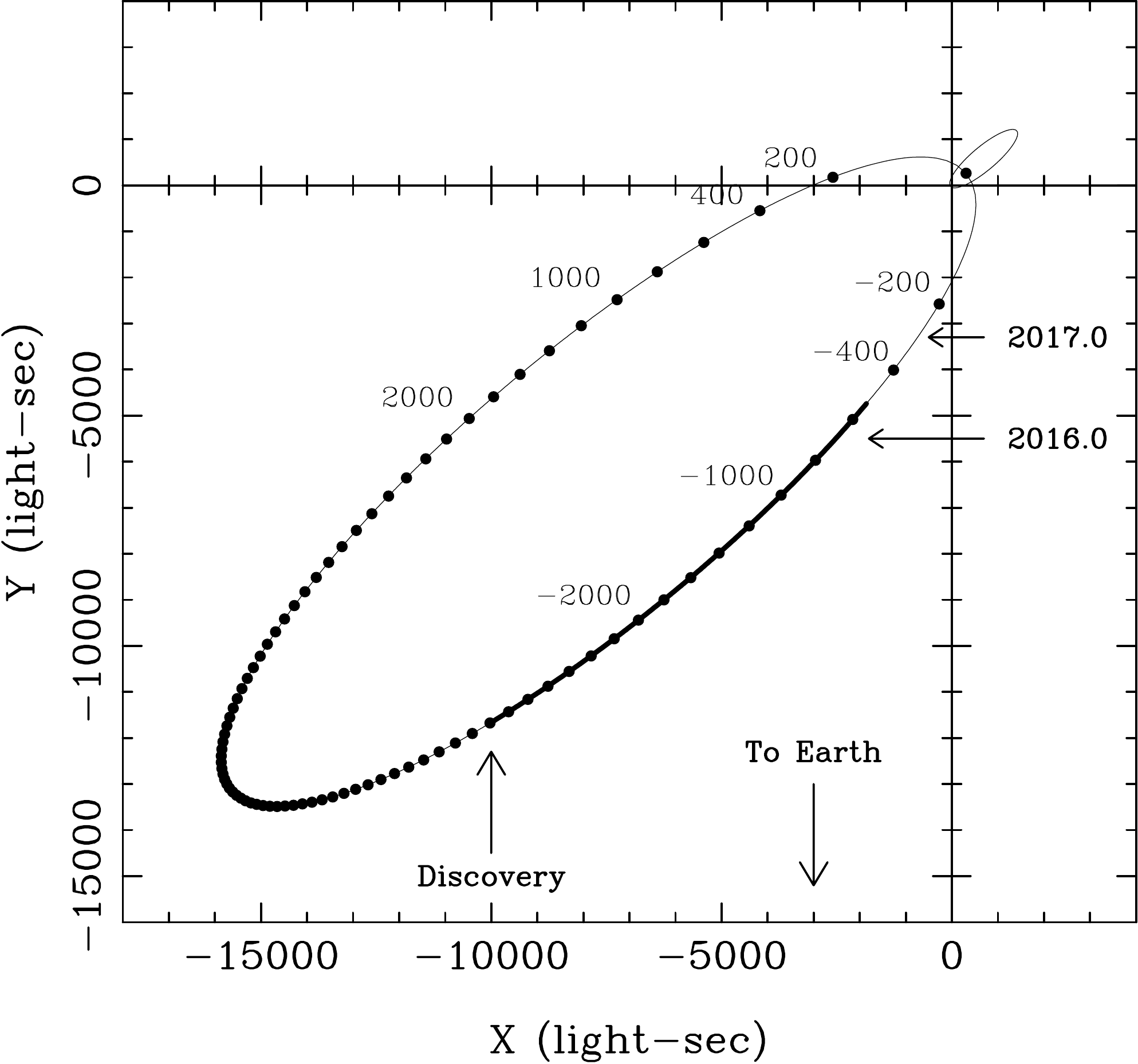}
\caption{A schematic illustration of the orbital motion of \psr\ and its Be-star companion. 
 Periastron is anticipated to take place in November 2017 \cite{Ho}. 
\label{fig:orbit}}
\end{figure}

Observations of the system have already shown a dramatic increase in the X-ray flux, increasing by a factor of about 70 since 2002, and about 20 since 2010 \cite{Ho}.
This suggests that interactions between \psr\ and the Be star winds have already begun.

\section{Comparison with Other Very High Energy \GR\ Binary Systems}
\GR\ binaries are a select class of sources (only six very high energy (VHE, E $>$ 100~GeV) \GR\ binaries have been detected \cite{TeVCat}) and the nature of the compact object is known in only one instance. 
Though the number of known \GR\ binaries is limited, comparisons with other systems containing Be stars shows the variety of potential outcomes of this periastron passage.
PSR~B1259-63/LS~2883 is the only system where the compact object is known (a radio pulsar) and is the most similar system to \psr/MT91~213.
As well as VHE emission detected by H.E.S.S \cite{B1259_HESS}, \fermi-LAT has also detected flaring activity about 30 days after periastron, with a peak flux about 20-30 times pre-periastron levels, almost reaching the estimated total spin down luminosity (\cref{fig:SwiftLC}) \cite{B1259_LAT}.
Observations of H.E.S.S~J0632+057 with VERITAS and \swift\ have shown variable and correlated VHE and X-ray emission \cite{VTS0632,VTS0632corr}.
It has recently been detected in high energy (HE, 0.1~GeV $<$ E $<$ 100~GeV) \GR s using \LAT\ data by Malyshev and Chernyakova \cite{Malyshev} and is in the 3FHL catalog (3FHL J0632.7+0550) \cite{3FHL}.
LS~I~+61~303 has been shown to undergo significant orbital and superorbital variability in both the HE and VHE bands \cite{VTSLSI}.
Most notably, all three of these sources show significant differences in their multiwavelength characteristics, in particular the relationship between the HE and VHE \GR\ emission and their dependencies on orbital phase.
If \psr/MT91~213 is a \GR\ binary it has the potential to provide significant insight into this rare class of systems.

\section{\swift\ Observations and Results}
\begin{figure}[tb]
\centering
\includegraphics[width=0.6\textwidth]{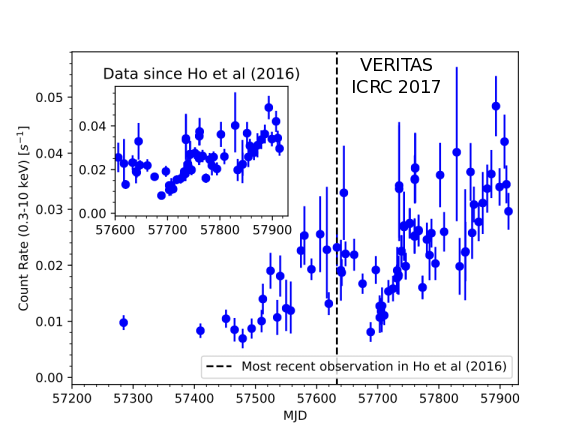}
\caption{\swift-XRT light curves of \psr\ binned by observation \cite{Leicester1,Leicester2}, a closer view of the data since Ho et al. \cite{Ho} is shown in the inset.}
 \label{fig:SwiftLC}
\end{figure}

\swift\ is an X-ray telescope designed to detect and study \GR\ bursts.
As a result it is a flexible, rapid-response instrument capable of performing a wider variety of observations.
In this work we use the X-Ray Telescope (XRT) which is sensitive to 0.2~--~10~keV X-rays (though we use a slightly narrower 0.3~--~10~keV band).
Ho et al. \cite{Ho} reported on \swift\ monitoring of \psr\ up to MJD 57633.16 (2016 September 2).
Since then monitoring has continued with observations taken every few days.
Light curves have been produced of the count rate binned by observation (\cref{fig:SwiftLC}).
They show that the X-ray count rate is continuing to rise, reaching a factor of five higher than the average level recorded in 2015 and and a factor of 35 higher than the average level recorded in 2013 (not shown in \cref{fig:SwiftLC}).
There is also evidence of variability in the light curve with peaks visible around MJD 57640 and 57860, the origin of this variability is uncertain.

\section{\LAT\ Observations and Results}
The \textit{Fermi Gamma-ray Space Telescope} has been operating since its launch in 2008 with two \GR\ instruments: the Large Area Telescope (LAT) and the Gamma-ray Burst Monitor (GBM). 
The LAT \cite{LAT}, \fermi 's primary instrument, is a pair conversion \GR\ detector that is sensitive to high energy (HE) \GR s with energies from 20~MeV to greater than 500~GeV. 

We have undertaken an analysis of \LAT\ Pass 8 data \cite{Pass8} using ``SOURCE'' class photons.
A joint likelihood analysis was performed with the events separated into event types FRONT and BACK.
Two different energy ranges were tested, 300~MeV to 300~GeV and a higher energy threshold of 1~GeV to 300~GeV to reduce the pulsar contribution.
The region of interest was a square of width 20\degree\ and the time interval covered one year, from MJD 57533 to 57897 (2016-05-24 to 2017-05-24).
A base model derived from the 3FGL catalog was used, with the \fermipy\ tool \emph{find\_sources} used to identify new sources. 
Any model components with a test statistic less than 25 were removed from the analysis and spectral models were updated if necessary.
The data were then binned into twelve bins and a light curve was produced.
Due to the variable nature of Cygnus~X-3, a known HE \GR\ emitter which lies 0.50\degree\ from \psr\ and was not detected in the analysis, a source was added in its location.
The data were then re-fit and another light curve produced to test whether any variation detected in \psr\ was a result of variation in the Cygnus~X-3 flux.

\begin{figure}[tb]
\centering
\begin{subfigure}[b]{0.48\textwidth}
\includegraphics[width=\textwidth]{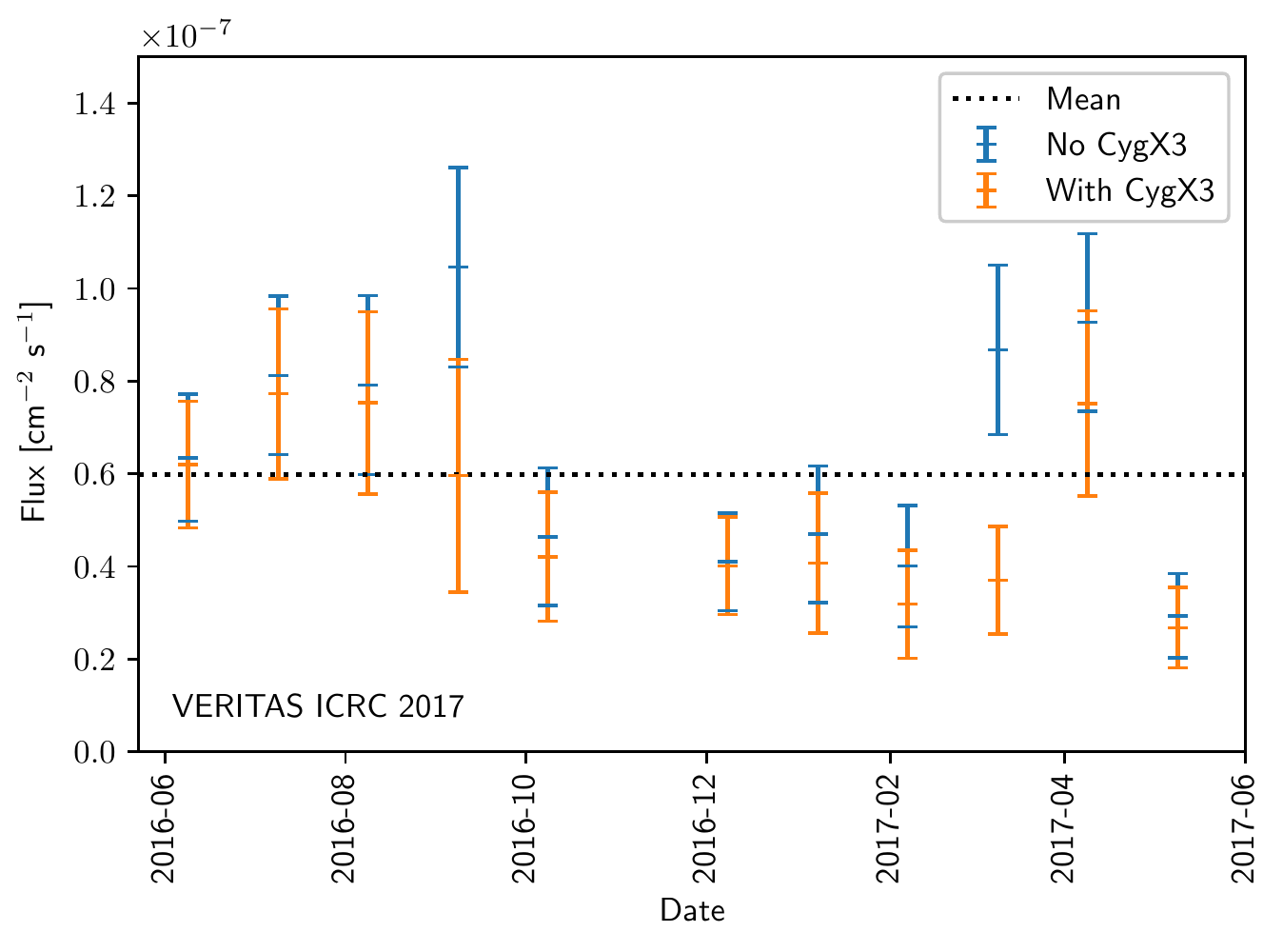}
\caption{300~MeV -- 300 GeV}
\end{subfigure}
\begin{subfigure}[b]{0.48\textwidth}
\includegraphics[width=\textwidth]{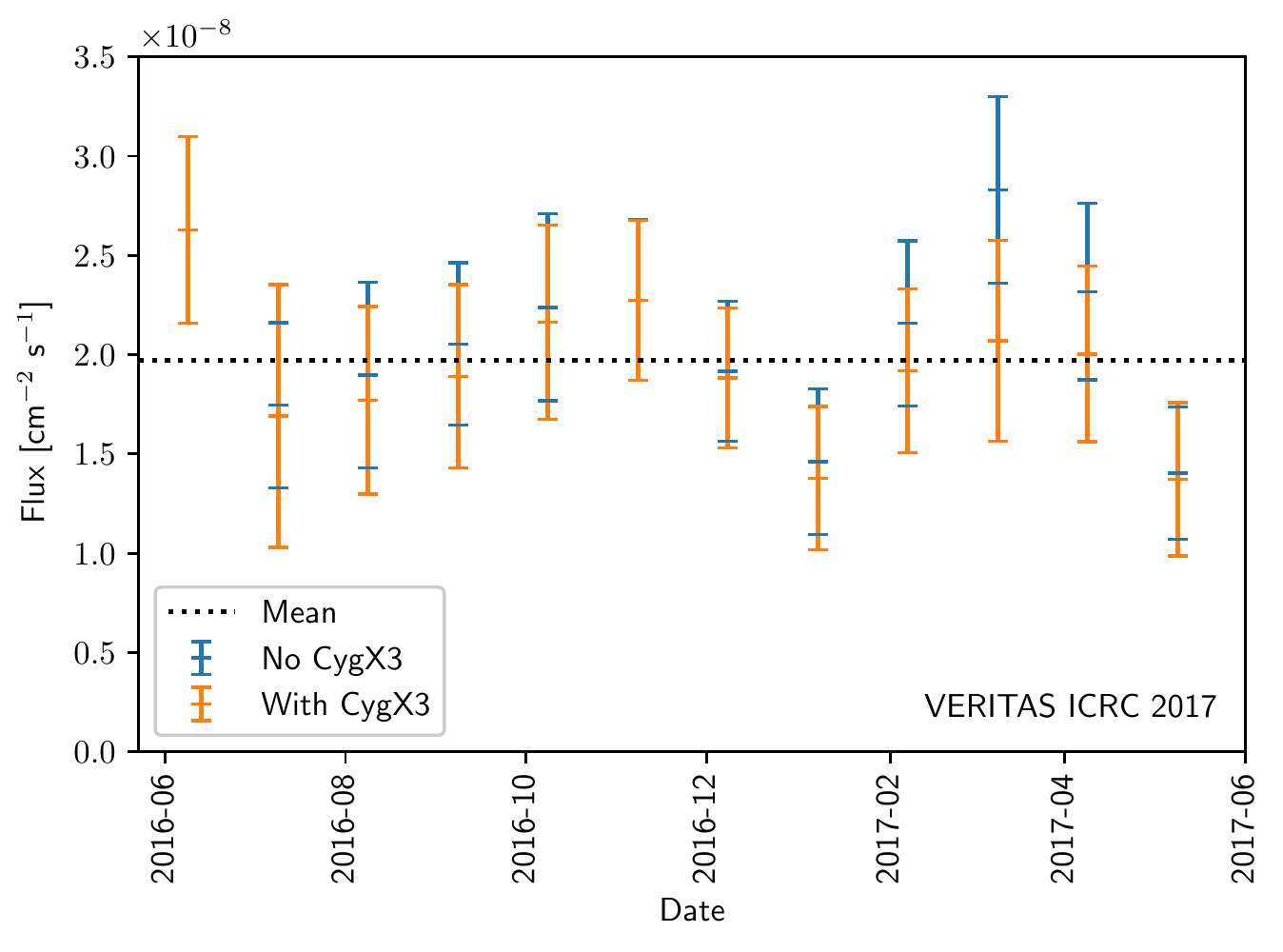}
\caption{1 -- 300 GeV}
\end{subfigure}
\caption{\LAT\ light curves covering the period MJD 57533 to 57897 for \psr.  Two background models were tested, one with Cygnus~X-3 (orange) and (because it was not significantly detected in the analysis) one without (blue).  The average flux over the period of the light curve (black dotted line) is also shown.
In all cases the variation is not significant with $\chi^2$/degrees of freedom of less than one.
\label{fig:FermiLC}}
\end{figure}

\psr\ is detected strongly in both analyses, with test statistics of 1003 and 817, and fluxes of (6.00\plm 0.40)~$\times 10^{-9}$ cm$^{-2}$~s$^{-1}$ and (2.00\plm 0.12)~$\times 10^{-9}$ for the analysis in the energy ranges 300~MeV~--~300~GeV and 1~GeV~--~300~GeV respectively.
The resultant light curves (\cref{fig:FermiLC}) show no marked increase, as is seen in the \swift\ data.
They show minor variation around the mean flux (taken from the fit with Cygnus~X-3 in the model) but, in comparison to a uniform flux model, have $\chi^2$ per degree of freedom less than one.
The introduction of Cygnus~X-3 noticeably reduces the flux for the 300~MeV~--~1~GeV data point at MJD 57836 (2017-03-24).
This is likely due to flaring activity from Cygnus~X-3 but will require further study to confirm this.

A steady state analysis was also conducted using the whole \LAT\ dataset up to MJD 57897.
The measured fluxes from \psr\ were (6.31\plm 0.13)~$\times 10^{-9}$ cm$^{-2}$~s$^{-1}$ and (2.67\plm 0.40)~$\times 10^{-9}$ cm$^{-2}$~s$^{-1}$ over the energy ranges 300~MeV~--~300~GeV and 1~GeV~--~300~GeV respectively.
They shows that the fluxes measured from MJD 57533 to 57897 have not increased relative to the average fluxes up to MJD 57897.

\section{\veritas\ Observations and Results}
The Very Energetic Radiation Imaging Telescope Array System (\veritas) \cite{veritas} is an array of four IACTs, located at the Fred Lawrence Whipple Observatory in southern Arizona (31\degree 40'~N, 110\degree 57'~W, 1.3~km a.s.l.). 
Each telescope is of Davies-Cotton design, with a 12-m diameter reflector comprised of 345 hexagonal mirror facets. 
The focal length of each telescope is 12~m and each telescope is equipped with a camera consisting of 499 close-packed photomultiplier tube (PMT) ``pixels'' at the focus.
The angular spacing between PMTs is 0.15\degree, which yields a total field of view (FoV) of 3.5\degree.
Full array operations began in 2007 and, in the summer of 2009, the first telescope was relocated to increase the sensitivity of the array \cite{T1}.
Following a trigger upgrade, in summer 2012, the cameras in each telescope were replaced with new, high quantum efficiency PMTs which has resulted in a decrease of the array energy threshold to about 85~GeV \cite{VERPerf}. 
All the data presented in this work were taken after the 2012 upgrade and using standard \veritas\ operational procedures. 
The results presented here were generated using one of the standard \veritas\ event reconstruction packages, similar to the scheme described in \cite{LSI} using a 0.1\degree\ integration radius (tuned for point sources).

\psr\ lies at the edge of the extended source \TeVJ\ (9.5\plm 1.2' along the major axis and 4.0'\plm 0.5' minor axis).  
In this work we did not study the emission from the extended source.
Rather we have focused on the emission from the direction of \psr, performing a point source search at the location of \psr.
We expect to have some level of contamination from \TeVJ\ within the analysis, however, changes in the flux will be indicative that at least part of the emission detected is from the binary system, as it is highly unlikely that the we will detect variability from \TeVJ\ due to its extended nature.

\begin{figure}[tb]
\centering
\includegraphics[width=0.6\textwidth]{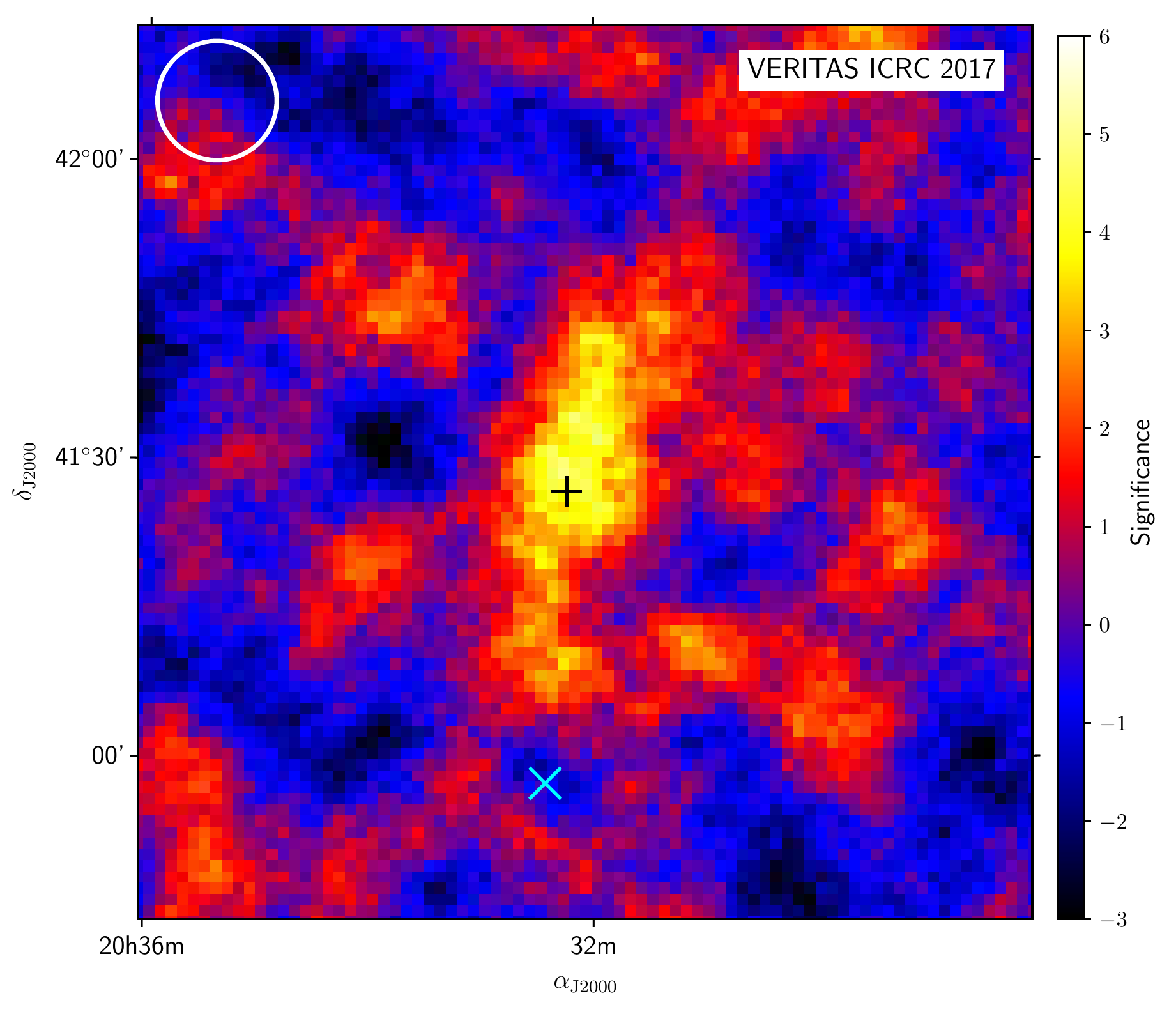}
\caption{\veritas\ significance sky map.  The location of \psr\ is shown with black "+", Cygnus~X-3 is shown with a cyan "$\times$".  The integration radius (which is equal to the PSF) is shown with the white circle in the upper left corner.
\label{fig:VerSigSky}}
\end{figure}

45.0 hours of data have been taken on \psr\ since the 2012 upgrade, giving 37.1 hours of quality-selected live time.
Of  this, 17.6 hours were taken in fall 2012, 8.5 hours in fall 2016 and 11.0 hours taken in spring 2017.
Further data are being taken in spring 2017 and data will also be taken in fall 2017.
Using these data, \GR\ emission is detected at the 4.82$\sigma$ level from the direction of \psr.
The peak significance is 5.69$\sigma$ (4.98$\sigma$ after accounting for the number of statistical trials) at (\ra, \dec) = (\hms{20}{32}{17}, \dms{41}{28}{30}), 0.03\degree away from the location of \psr\ and well within the point spread function (PSF) which is 0.1\degree\ at 1~TeV.
A significance sky map for the region is shown in \cref{fig:VerSigSky}.

To test for variation in the flux with time, the data was broken up into yearly bins.
The results are presented in \cref{tab:VEGASUL}.
None of the yearly datasets reveal a more than a marginally significant detection. 
We also note that much of the 2012 data were taken at a large offset from the location of PSR J2032+4127. 
There is no change in the integral flux measured within the large statistical errors on the measured values.  

\begin{table}
\centering
\caption{The results of the \veritas\ observations broken up by year of observation. 
All of the results are from the RR analysis.
The significance (Sig) is given in unit of $\sigma$, the energy threshold (E$_{Thresh}$) is given in TeV and the Flux is integral flux in units of $10^{-13}$~cm$^{-2}$~s$^{-1}$ above the energy threshold.
\label{tab:VEGASUL}}
\vspace{1mm}
\begin{tabular}{cccccccccc}
\hline
Year& Live Time [hr] & Sig & Sig/$\sqrt{\mathrm{hr}}$ & E$_{Thresh}$ & Flux    \\ \hline
2012 & 17.62 & 2.59 & 0.78 & 0.45 & 2.9\plm3.8 \\
2016 & 8.49 & 3.29 & 0.99 & 0.45 & 7.5\plm3.1 \\
2017 & 11.0 & 4.17 & 1.25 & 0.45 & 7.0\plm2.9 \\ \hline
\end{tabular}
\end{table}

Given the morphology reported in \cite{VTS_TeV2032}, we would expect to measure a flux that is about 1/5 of the flux they report, assuming that it smoothly follows their reported two dimensional Gaussian morphology.
With a differential flux normalization of (9.5\plm 1.6$_{stat}$\plm 2.2$_{sys}$)$\times 10^{-13}$~TeV${-1}$~cm$^{-2}$~s$^{-1}$ at 1~TeV and a spectral index of 2.10\plm 0.14$_{stat}$\plm 0.21$_{sys}$.
This gives a predicted integral flux above 0.45~TeV of about (4.4\plm 1.0$_{stat}$)$\times 10^{-13}$ for this analysis, below, but within statistical errors of, the flux measured in this work.
At present there is no evidence that the emission observed is related due to the binary system, further observations are required to investigate this in more detail.

\section{Future Plans}
With \psr\ approaching periastron, the observation campaign is only at the beginning.
The X-ray flux shows significant variation and this suggests that \psr\ is beginning to interact with the stellar wind of MT91~213 and thus the possibility of detecting HE and VHE \GR s from the binary interaction is increasing.
\veritas\ will be conducting targeted observations in the fall of 2017 to search for enhanced VHE emission.

As a survey instrument, \LAT\ exposure will continue, providing a continuous monitor of \psr\. 
They are also considering modifying the observing strategy to increase the sensitivity to variability in the HE waveband.
A dedicated analysis is being conducted by members of the \fermi\ collaboration which will search for enhanced emission from \psr\ and will include an off-pulse analysis search for changes in the continuous emission from the region.

\swift\ observations are also ongoing, and we will continue to monitor the X-ray flux through periastron.
\swift\ also has an ultraviolet/optical telescope (UVOT) which is co-aligned with the XRT.
Data from this will also be analyzed to search for variation in these wavelengths.

Using these three datasets (and other observations if available), a search will be conducted for correlated emission between the wavebands in the hope that this will identify a new \GR\ binary.

\section{Conclusions}
\psr\ is a \LAT\ pulsar which is believed to be in a long period binary system with the Be star MT91~213.
Approaching periastron (expected in November 2017) it is already showing significant variability in the X-ray flux, with count rates reaching a factor of 35 times the level recorded in 2013 and also evidence of variability within the observed rise.
A light curve of one year of \LAT\ data shows no evidence for a change in flux using monthly binning or by comparing the flux from the last year with that from the whole mission.
37.1 hours of \veritas\ data have been examined to search for point source emission from the direction of \psr, with \GR\ emission detected at the 4.82$\sigma$ level.
Located within \TeVJ, an estimate was made on the expected flux from the extended source at (4.4\plm 1.0$_{stat}$)$\times 10^{-12}$~cm$^{-2}$~s$^{-1}$, this is below the flux measured in this work, but within statistical errors.
There is no evidence of emission from the binary system at this stage.
Work is ongoing to analyze additional archival data and future observations will be conducted to search for variation in the VHE flux, which will be indicative of VHE emission from the \psr/MT91~213 system.

\section{Acknowledgements}
This research is supported by grants from the U.S. Department of Energy Office of Science, the U.S. National Science Foundation and the Smithsonian Institution, and by NSERC in Canada. We acknowledge the excellent work of the technical support staff at the Fred Lawrence Whipple Observatory and at the collaborating institutions in the construction and operation of the instrument. The VERITAS Collaboration is grateful to Trevor Weekes for his seminal contributions and leadership in the field of VHE gamma-ray astrophysics, which made this study possible.

This work made use of data supplied by the UK Swift Science Data Centre at the University of Leicester.

\end{document}